\documentclass[amsmath,amssymb,lengthcheck,onecolumn,aps,prl] {revtex4}
\usepackage[T1]{fontenc}
\usepackage[utf8]{inputenc}
\usepackage{graphics}
\usepackage{graphicx}

\usepackage{xcolor}

\begin{document}
\title{\textbf{Surface excitation of Rydberg dressed quantum droplet of Bose-Einstein Condensates}}
\author{Avra Banerjee}

 \author{Dwipesh Majumder}
\affiliation{Department of Physics, Indian Institute of Engineering Science and Technology, Shibpur, W.B., India}
\date{\today}

\begin{abstract}

We have considered a quantum droplet of two components of Bose-Einstein condensate (BEC) inside the electron of a Rydberg atom to study the surface mode of collective excitation using the Bogoliubov theory of excitation. We have calculated the surface excitation spectrum for various Rydberg electron-atom interaction strengths. From the energy spectrum, we calculated the surface tension of the droplet as a function of Rydberg electron-atom interaction strength. Our study shows that the electron-atom interaction enhances the surface energy; hence, the droplet will be more stable inside the electron of a Rydberg atom.
\end{abstract}

\maketitle

\section*{Introduntion}

A new paradigm for investigating the fascinating quantum many-body physics has emerged as a result of the theoretical prediction \cite{Petrov2015} and practical demonstration of an ultra-dilute quantum droplet \cite{LHY1,LHY2,Trarruell2018,expt_review} containing cold atoms. This new state of matter is based on the balance between the repelling force originating from beyond-mean-field quantum fluctuations \cite{Petrov2015} and the mean-field attraction.
 The repulsion due to the quantum fluctuation balances the mean-field attraction and stabilizes the system against collapse because of its sharper density scaling. Experimental study was done on the quantum droplets of a mixture of two species of alkali Bose atoms. In addition to this system, the quantum droplet has also been observed as a cigar-shaped droplet in the asymmetric dipolar interacting systems \cite{drop_exp2}. Using the extended Gross-Pitaevskii equation (GPE) with the LHY energy functional $E_{LHY}$, the experimental results can be qualitatively interpreted \cite{tpau}.

To keep things simple, we adhere to Petrov's \cite{Petrov2015} original proposal and take into account a binary Bose mixture with attractive interspecies interactions \cite{avra_PT}, in which a spherical self-bound droplet has been observed \cite{droplet2}. Contrarily, the anisotropic and cigar-shaped geometry of the dipolar quantum droplet disfavor the surface modes \cite{surface1,surface2,surface3}. 

With an emphasis on the practical viability of observing the surface modes with exotic dispersion relationships, our goal in this work is to theoretically understand the collective excitations of a three-dimensional spherical ultra-dilute quantum droplet inside the electron of a Rydberg atom.  Bose-Einstein condensation (BEC) of ultracold atoms confined in a Rydberg atom \cite{Polaron_The2016,Polaron_Exp2018,Polaron_3}, known as Rydberg polaron, is a fascinating research topic. The first Rydberg atom in ultracold atomic BEC was proposed by Greene et al. in 2000 \cite{PRL85.2458}. Many aspects of the Rydberg atom in BEC have been investigated so far \cite{QComputer,PRL85.2458,avra_ryd}. Here, we are listing some of them for completeness. There are some studies looking at the existence of impurities (ion or Rydberg atom) in the BECs at ultracold temperatures \cite{ryd1,ryd2}. In-depth research has been done on cutting-edge methods for manipulating ion-atom combinations, including cold collision, chemical reactions \cite{cold_chem1, cold_chem2, cold_chem3, cold_chem4}, and single ions in BEC \cite{single_ion}. 
Phase variation brought on by the Rydberg atom's motion was used to monitor the Rydberg atom in the BEC and study its motion.
We may comprehend the motion of the Rydberg atom in the mixture using the relevant phase data  \cite{ryd_dy6,GP_Ry2}. The impurity-based BEC may be utilized to study micro-macro and macro-micro entanglement, to list a few examples \cite{entan1,entan7}. Additionally, BEC has complete control over the micro-macro quantum system \cite{mic_mac}. Nowadays, researchers routinely create giant Rydberg atoms with principal quantum number around $n_r \ge 300 $ \cite{ryd_300}, which corresponds to an orbital radius of several micrometers, which can confine thousands of Bose atoms at ultra-cold temperatures.

Our system is finite and spherical-symmetric; the energy spectra of the quantum droplet are given by $\omega_{ln}$, where $l$ and $n$ are the droplet's angular momentum quantum number and the droplet's radial quantum number, respectively. By using the LHY energy function \cite{lhy3,lhy4,lhy5}, the excitation spectrum of a spherical droplet, including the lowest monopole mode (i.e., the breathing mode with $l = n = 0$) and surface modes (i.e., $l \geq 2$ and $n = 0$), has already been addressed \cite{hui}. For nonzero angular momentum $l \neq 0$, the radial quantum number $n$ denotes the number of nodes in the radial wave functions. However, at zero angular momentum, the nodeless wave function is the condensate wave function and is excluded as a wave function of Bogoliubov quasiparticles. In the $l = 0$ sector, therefore, the number of nodes in the radial wave functions is given by $n + 1$. Hence, the breathing mode has a node in its radial wave function \cite{hui}.

We have studied here how the droplet's surface mode of excitation ($ \omega_{ln} $; $l\geq 2$ and $n=0$) changes in the presence of the electron of the Rydberg atom. Surface mode of excitation has been observed in several experiments \cite{surface_mode_exp,surface_mode_exp1}. We have considered that there is no external trapping potential. However, maintaining a finite harmonic trapping potential might increase the quantum droplet's stability. In this regard, it is interesting to think about the collective excitations of the droplet in the presence of an external trapping potential, in particular, the existence of the surface modes \cite{hui,surface_ex}. The dipolar mode $(l =1)$ can be regarded as a center-of-mass displacement of the droplet; therefore, $\omega_{10} =0 $ \cite{Petrov2015,hui,pu}.
We point out that Cikojevi and his colleagues \cite{QMC} have most recently examined the breathing mode $\omega_{00 }$ and quadrupole mode $\omega_{20 }$ of a self-bound spherical $^{39}$K droplet utilizing the time-dependent extended GPE equation \cite{Ferioli} and the precise Diffusion Monte Carlo (DMC) energy functional.

Despite having finite sizes, the liquid droplets have entirely different characteristics from confined BEC. Except near the surface, the density inside the droplet is constant. When droplet forms inside the electron of a Rydberg atom, there is an interaction of the Rydberg atom's electron with the surface of the droplet. Here, we have studied how surface excitation changes with the various interaction strengths of the Rydberg atom's electron. We have used the exact numerical solution of the Bogoliubov equations resulting from the linearized time-dependent extended Gross-Pitaevskii equation.


\section*{Model and calculation}

We consider a mixture of two species of Bose atoms (two different internal degrees of freedom of the same isotope of an element, ($m_1=m_2=m$, where $m$ is the mass of the condensed atom) inside a Rydberg electron ($\mathcal{V}_0 = 2\pi \hbar^2 a_e/m_e$ is the Rydberg electron-atom interaction strength, where $a_e$ is the electron atom scattering length and $m_e$ is the mass of an electron) \cite{s_wave,jia,shiv}.  In the ultracold temperature and very low density, the interaction potential between two neutral atoms can be written as $V(\vec{r}_1,\vec{r}_2) = {g_{ij}} \delta(\vec{r}_1-\vec{r}_2)$, where ${g_{ij}} = 4\pi \hbar^2 a_{ij}/m$ is the strength of the interaction potential between the atoms of $i^{th}$ and $j^{th}$ species, $a_{ij}$ is the s-wave scattering length, which can be controlled by magnetic Feshbach resonance. The mean-field GP equations are not sufficient to
study the nature of the condensate; we need to consider self-repulsive beyond-mean-field, Lee-Huang-Yang (LHY) \cite{LHY_review,length} term (quantum fluctuation term, $g_{LHY}$ is the strength of the quantum fluctuation). The coupled non-linear GP equations \cite{Petrov2015, GP_eqn} with the LHY interaction and Rydberg dressing in two species can be written as


\begin{small}
\begin{eqnarray}
 i\hbar\frac{\partial \phi_1(r,\tau)}{\partial \tau}= &  \Big[-\frac{\hbar^2\nabla  ^2}{2m}+g_{11}|\phi_1|^2-g_{12}|\phi_2|^2 +  g_{LHY}|\phi_1|^3  \nonumber \\
 & +\mathcal{V}_0  |\Psi_R(r)  |^2\Big]\phi_1(r,\tau)
  \label{gp1}
  \end{eqnarray}
  \begin{eqnarray}
 i\hbar\frac{\partial \phi_2(r,\tau)}{\partial \tau}= & \Big[-\frac{\hbar^2\nabla  ^2}{2m}+g_{22}|\phi_2|^2-g_{12}|\phi_1|^2 +g_{LHY}|\phi_2|^3     \nonumber \\
 & +\mathcal{V}_0  |\Psi_R(r)  |^2\Big]\phi_2(r,\tau)
  \label{gp2}
\end{eqnarray}
\end{small}


where the first term on the right side is the kinetic energy term, the second term is intraparticle repulsive interaction, the third term is interparticle attractive interaction (the value of $g_{12}$ is positive), the fourth cubic term is responsible for quantum fluctuation, and the last term is the interaction between atom and Rydberg electron. In our numerical calculations, we have used the length in unit of $l_{0}$ (defined in equation (9) of ref. \cite{Petrov2015}), time in $\frac{ml_0^2}{\hbar}$, energy in $\frac{\hbar^{2}}{ml_{0}^{2}}$ unit. The Rydberg atom wave function, $\Psi_R(r) = \Psi_{n_r 00}$ \cite{s_wave}, is the hydrogen atom wave function with large $n_r$ (here we have used symbol $n_r$ for principle quantum number to avoid confusion with the excitation mode of the condensate). We have used single-mode approximation of the coupled equations (\ref{gp1} and \ref{gp2}) to get equation (\ref{single_mode}), and equation (\ref{single_mode}) is expressed in scaled form according to ref. \cite{adhikary_dless}. The mean-field interaction strength $\delta g= -g_{12}+\sqrt{g_{11}g_{22}}$. We have considered intra-species coupling constants ($g_{11}=g_{22}=g$). Equal number of particles have been taken into account for each species ($N_1=N_2=N/2$), where $N_1$ and $N_2$ are the number of particles of the first and second species, respectively. The GP equation in scaled form \cite{Petrov2015,adhikary_dless} can be written as


\begin{eqnarray}
  i \frac{\partial \phi(r,\tau)}{\partial\tau}&=& \Big[-\frac{\nabla^2}{2}-3 |\phi|^2+\frac{5}{2}|\phi|^3 \nonumber \\
  &&~~~+V_0|\Psi_R(r)|^2\Big]\phi(r,\tau)
  \label{single_mode}
\end{eqnarray}
  The condensate wave function follows the normalization condition,
\begin{equation}
  \int d^3r | \phi(r,\tau) |^2= N,
\end{equation}
$N$ is the total number of atoms in the droplet.
We have used the imaginary-time split-step Crank Nicolson method \cite{Adhikary} to solve the GP equation (\ref{single_mode}). One can find the ground state wave function and, hence, the ground state energy and chemical potential.

The excitations of the system can be obtained using the Bogoliubov method by considering the fluctuation over the ground state ($\phi_0$) as
$\phi(r,\tau) = \phi_0(r) e^{-i\mu \tau}+\delta \phi(r,\tau)$. The perturbation part can be written as 

\begin{equation}
  \delta \phi = \sum_j\left (u _j(\textbf{r})e^{ -i(\mu + \omega_j) \tau} + v_j^*(\textbf{r}) e^{-i(\mu-\omega_j) \tau}\right )
\end{equation}


Putting this in the GPE (equation (\ref{single_mode})) and linearizing with $\delta \phi$ we get the equation for the expansion coefficients as

  \begin{eqnarray}
\left(
    \begin{tabular}{c c c c}
      $\mathcal{H}+\mathcal{M}$ & $\mathcal{M}$\\
       $\mathcal{M}$ & $\mathcal{H}+\mathcal{M}$ \\

    \end{tabular}
\right)
\left(
\begin{tabular}{c}
$u_j(\textbf{r})$ \\$v_j(\textbf{r})$
\end{tabular}
\right) 
= \omega_j 
\left(
\begin{tabular}{c}
$u_j(\textbf{r})$ \\$-v_j(\textbf{r})$ 
\end{tabular}
\right) 
\label{matrix_eq}
\end{eqnarray}

where we have defined the operators,
\begin{eqnarray}
\mathcal{H}&=&\mathcal{L}-\mu \nonumber \\
\mathcal{M}&=&-3\phi_0^2+\frac{15}{4}\phi_0^3 \nonumber
\end{eqnarray}
 operator $ \mathcal{L}$ and the chemical potential ($\mu$) of the system are 
\begin{small}
  \begin{eqnarray}
   \mathcal{L} &=&-\frac{\nabla^2}{2}-3\phi_0^2+\frac{5}{2}\phi_0^3+V_0|\Psi_R(r)|^2 \\
   \mu &=& \frac{4\pi}{N} \int dr \left [\frac{1}{2}|\vec \nabla \phi_0 |^2-3|\phi_0|^4+\frac{5}{2}|\phi_0|^5+V_0|\Psi_R|^2|\phi_0|^2 \right ] \nonumber 
 \end{eqnarray}
 \end{small}
 $  $


   \begin{figure}
   \includegraphics[width=0.45\textwidth]  {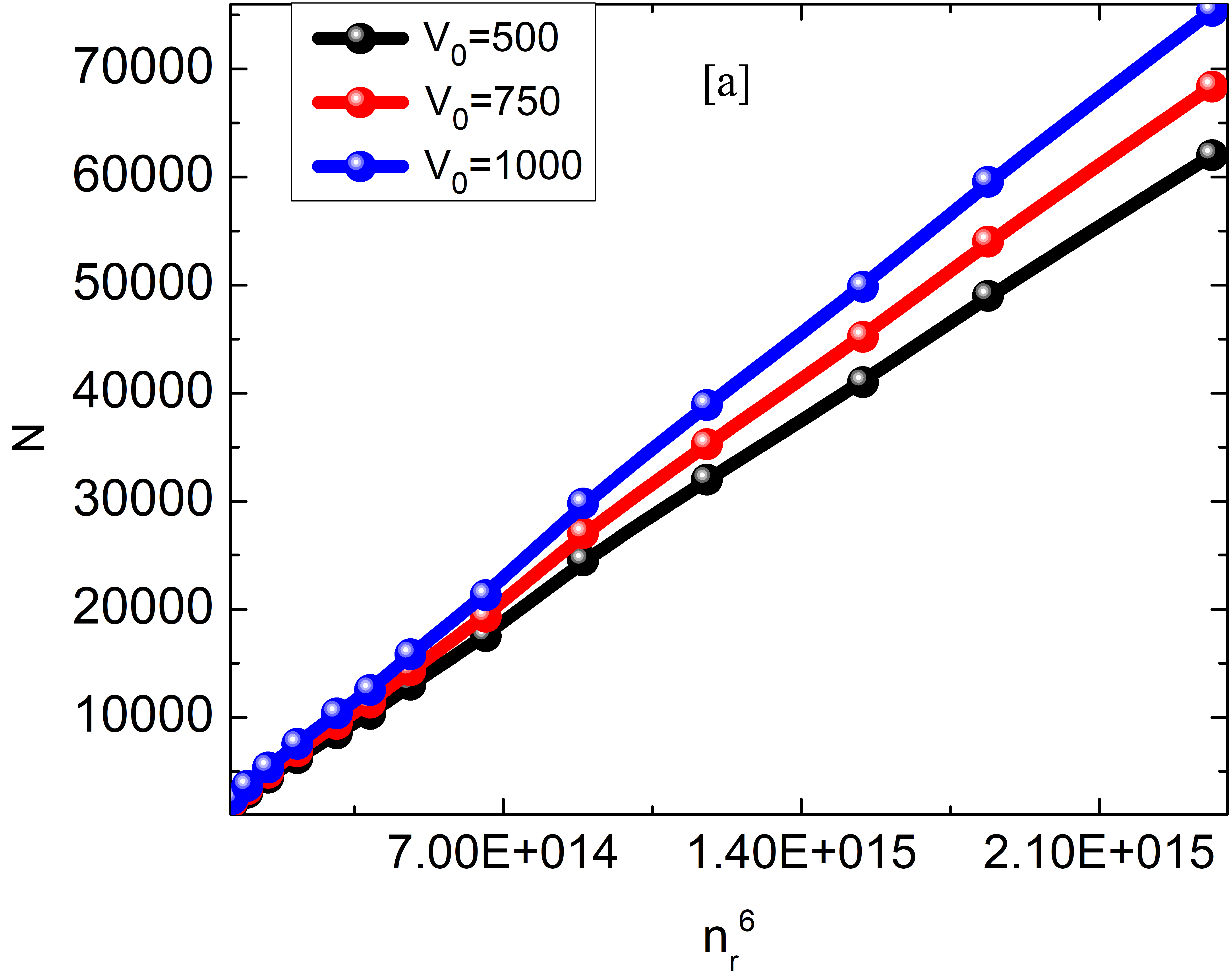}
 \includegraphics[width=0.45\textwidth]  {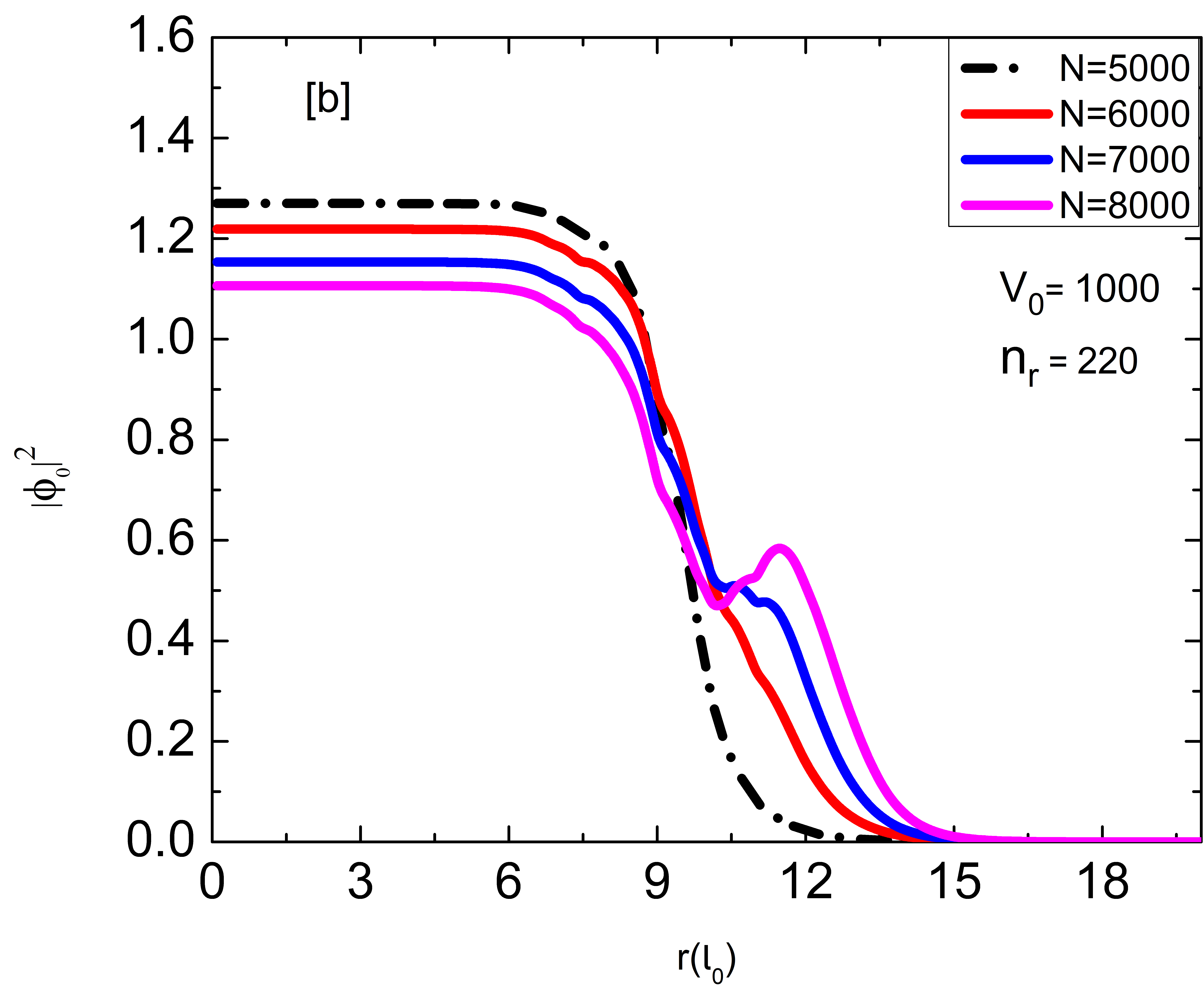}
     \caption {\textbf{[a]} The volume enclosed by the electron cloud is proportional to $n^6_r$, where $n_r$ is the principal quantum number of the Rydberg electron. We have calculated the maximum number of atoms ($N$) enclosed by the Rydberg atom's electron for different values of $n_r$ and the Rydberg electron-atom interaction strength $V_0$. It is seen that the maximum number of particles inside the Rydberg electron increases with the strength of the interaction $V_0$. Each plot is a straight line, and the density of the droplet is the same for each line (i.e., for a given $V_0$); the only difference is that the radial extension (R) of the droplet increases with a larger combination of  $N$ and $n_r$. \textbf{[b]} The maximum number of particles ($N$) inside the Rydberg electron is obtained by solving the GP equation (\ref{single_mode}) (see black dash-dotted curve). If the number of particles ($N$) increases further, the excess particles go beyond the electron.  There is a dip in the "pink curve" ($N = 8000$), which is the position of the electron of the Rydberg atom, and the peak in that curve is formed outside of the Rydberg atom.}
     
   \end{figure}

The Bogoliubov equations (equation (\ref{matrix_eq})) are coupled equations for $\{u_j, v_j\}$. It will be easy to get the energy eigenvalue $\omega_j$ if we could make the equations decouple. For that, we follow the work by Hutchinson, Zaremba, and Griffin \cite{hutch}, and introduce the auxiliary functions
\begin{equation}
\psi_j^{\pm}  \equiv u_j(r) \pm v_j(r)
\label{u_j}
 \end{equation}

And we have the decoupled equations of $\{\psi_j^{+}, \psi_j^{-}\}$
\begin{eqnarray}
    \mathcal{H}(\mathcal{H}+2\mathcal{M})\psi_j^{+} &=& \omega_j^2\psi_j^{+} \\
  ( \mathcal{H}+2\mathcal{M})  \mathcal{H} \psi_j^{-} &=&  \omega_j^2\psi_j^{-}
  \label{eq_psi_}
\end{eqnarray}

The two auxiliary functions are related to each other by
 \begin{equation}
  \mathcal{H}\psi_j^{-}=\omega_j\psi_j^{+}
\end{equation}
 
Our system is spherically symmetric; $ j$ must be a combination of two quantum numbers ($j = (ln)$): angular momentum and radial quantum number. One can use either equation for $\psi^-_j$ or equation for $\psi^+_j$ to calculate the excitation energy ($\omega_j$) of the system. Here, we have chosen the equation for $\psi_j^-$.


 \begin{figure}
   \includegraphics[width=0.52\textwidth]  {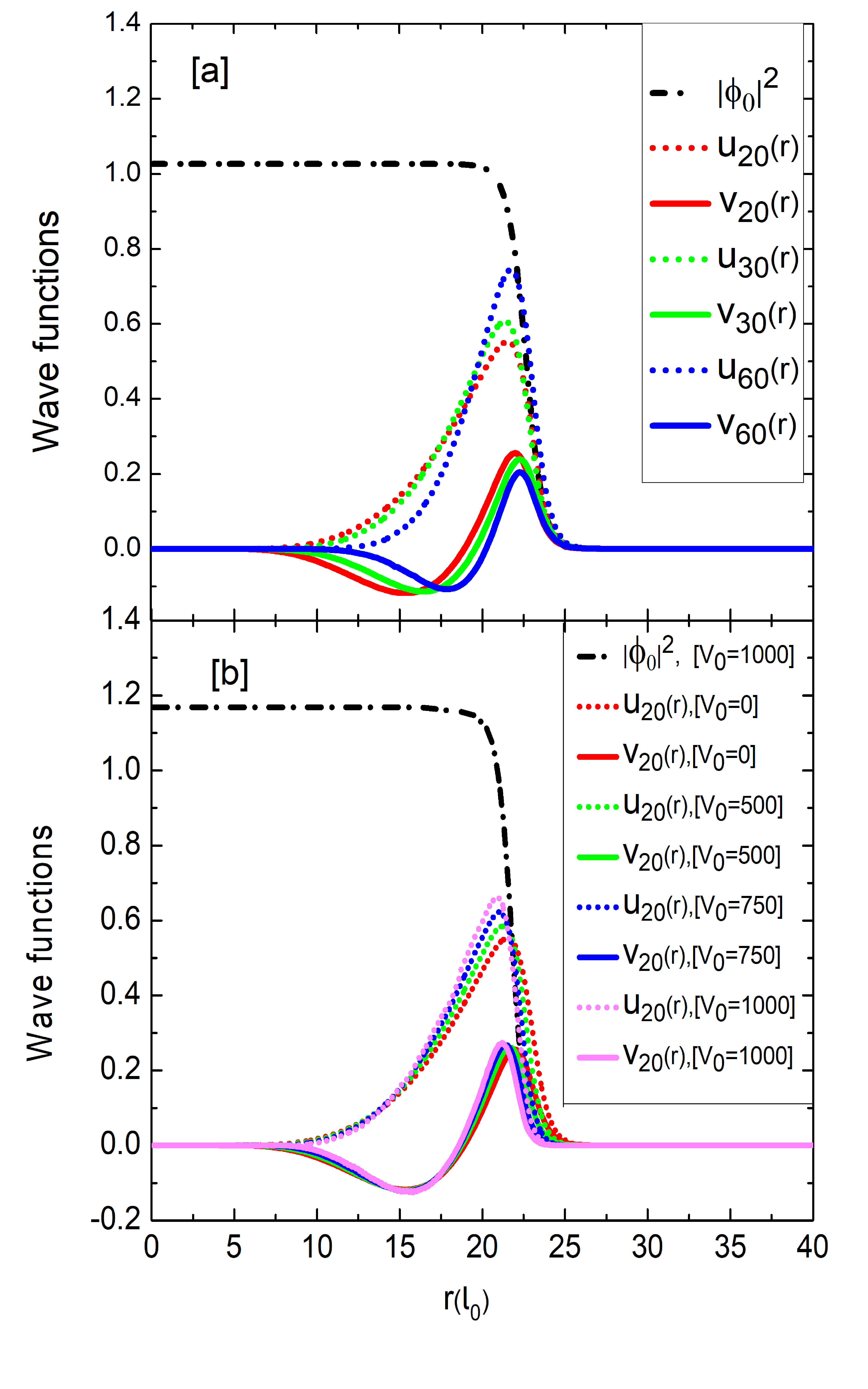}
     \caption{ Excited state wave functions. The black dash-dot line is the ground-state density of the droplet. The excited states are localized near the edge of the droplet. [a] excited state wave functions $u_{l,0}$ and $v_{l,0}$ of free droplet with particle number $N=50000$ for different $l$. [b] wave functions of Rydberg dressed quantum droplet. Here, we have plotted the wave function ($u_{20},\, v_{20}$) for various electron-atom interactions to see the Rydberg dressed effect on the excitation.}
\end{figure}

{\bf Expansion of $\psi_j^-$ on a suitable basis:} 
The proper basis function for the system can be obtained by solving the  equation
\begin{eqnarray}
  &&\mathcal{L} \Phi_{l\alpha} = \epsilon_{l\alpha} \Phi_{l\alpha} \nonumber \\
  \Rightarrow &&\mathcal{H} \Phi_{l\alpha} = \tilde{\epsilon}_{l\alpha} \Phi_{l\alpha}
\end{eqnarray}
here, $\alpha$ is the radial quantum number and $l$ is the angular momentum quantum number, and $\tilde{\epsilon}_{l\alpha} = \epsilon_{l\alpha} -\mu$.
For a given angular momentum $l$, we then expand the function $\psi_j^-$ (where $j=(ln)$) as
\begin{equation}
 \psi_j^{-}=\sum_\alpha^M c_\alpha^n \Phi_{l\alpha}(r) 
 \label{psi_j-}
\end{equation}
   in terms of the normalized eigenfunction basis $\Phi_{l\alpha}(r) $, where M is the number of states in the basis; in principle, M is infinite, but we have considered it finite but large to handle it numerically without compromising the accuracy of the calculation. 
Putting this wave function in equation (\ref{eq_psi_}) and using the orthogonality relation of $\{\Phi_{l\alpha}\}$, we obtain a secular equation \cite{hui}
\begin{equation}
\sum_\beta \left [\tilde{\epsilon}^2_{l\alpha} \delta_{\alpha\beta} 
  + \sqrt{\tilde{\epsilon}_{l\alpha} \tilde{\epsilon}_{l\beta}} \mathcal{M}^l_{\alpha\beta} \right ]
   \sqrt{\tilde{\epsilon}_{l\beta}} c_\beta^n
   =\omega_{ln}^2\sqrt{\tilde{\epsilon}_{l\alpha}}c_\alpha^n
\label{eq_eig_mode}
\end{equation}
where the matrix element is given by,
\begin{equation}
\mathcal{M}^l_{\alpha\beta} = 2\int d^3\vec{r} \; \Phi_{l\alpha}^* \mathcal{M} \Phi_{l\beta} 
\end{equation}


We have arranged the equation (\ref{eq_eig_mode}) in such a way that the secular matrix will be symmetric so that the numerical results will be more accurate. We have solved the equation using matrix diagonalization code from LAPACK \cite{lapack} for a given $l$ ($l$ ranging from $2$ to $9$ and $n = 0$.).

The excited state wave function $\{u_{ln}, \, v_{ln}\}$ can be obtained using equation (\ref{u_j}),(\ref{psi_j-}) and (\ref{v_j}).
\begin{equation}
\psi_j^{+}=\frac{1}{\omega_j}\sum_\alpha^M c_\alpha^n{\tilde{\epsilon}_{l\alpha}} \Phi_{l\alpha}(r)
\label{v_j}
\end{equation}

   \begin{figure}
   \includegraphics[width=0.50\textwidth]  {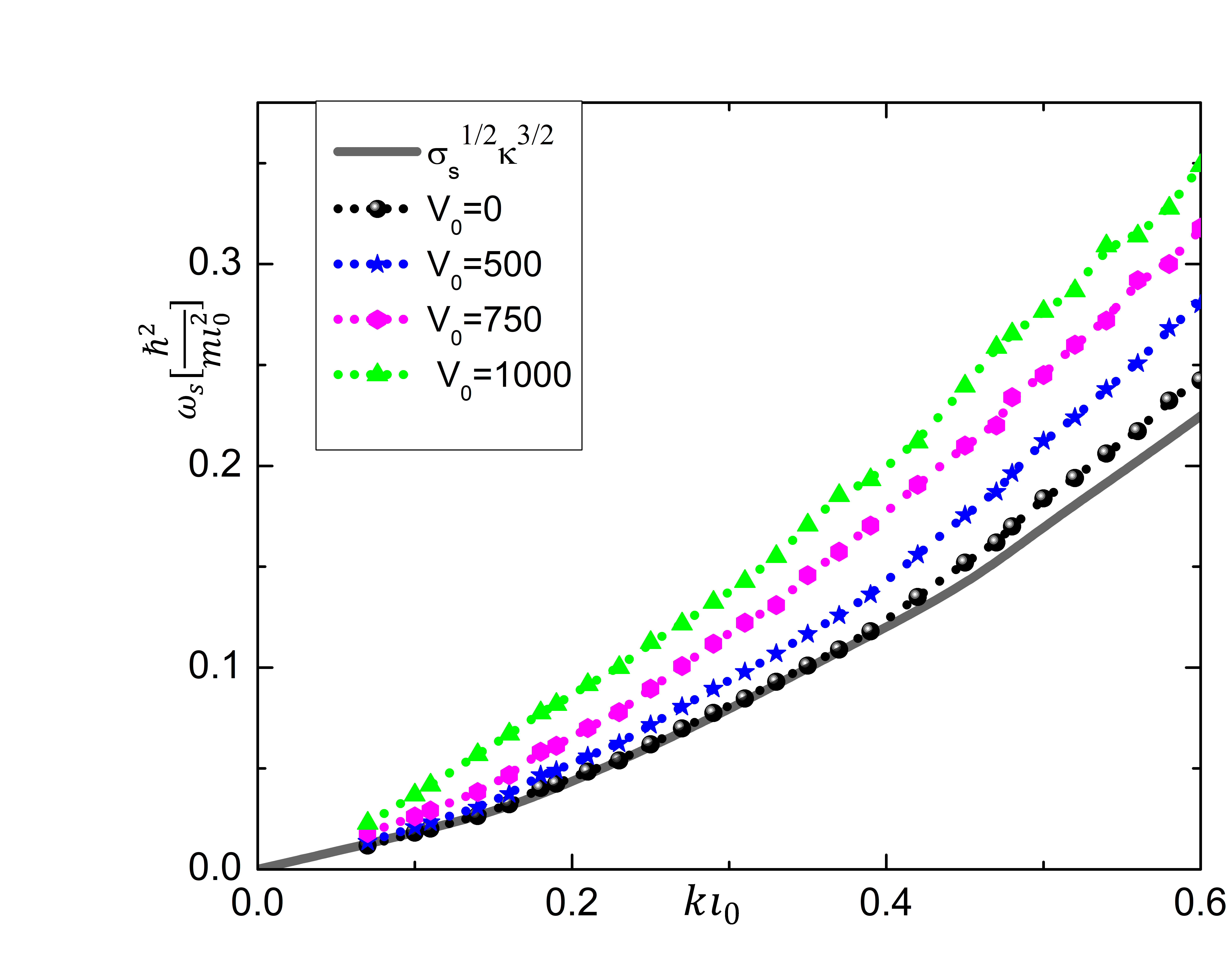}
     \caption{Excitation spectra ($\omega_s$) for different values of $V_0$ are obtained by diagonalizing the dispersion relation (\ref{eq_eig_mode}) with $n=0$. We have considered the number of particles ($N$) ranging from 2000 to 100000 to calculate the energy eigenvalue for angular momentum ranging from $l = 2$ to $l = 9$. We have plotted the energy as a function of an effective wave vector $k=[l(l-1)(l+2)]^{1/3}/R$ \cite{momentum}, where $R=\sqrt{5/3}r_{rms}$, where $r_{rms}$ is the rms radius of the droplet. Various values of $k$ can be obtained by changing $l$ and $R$, and $R$ depends upon the number of particles (N). The black solid line is the expected dispersion spectra (\ref{ripplons}) for surface modes ($\omega_s= \sqrt{\sigma_s} k^{3/2}$). The expected dispersion spectra match our data points for a lower effective wave vector. We have used these regions to calculate the surface tension.}
\end{figure}

\begin{figure}
      \includegraphics[width=0.45\textwidth]  {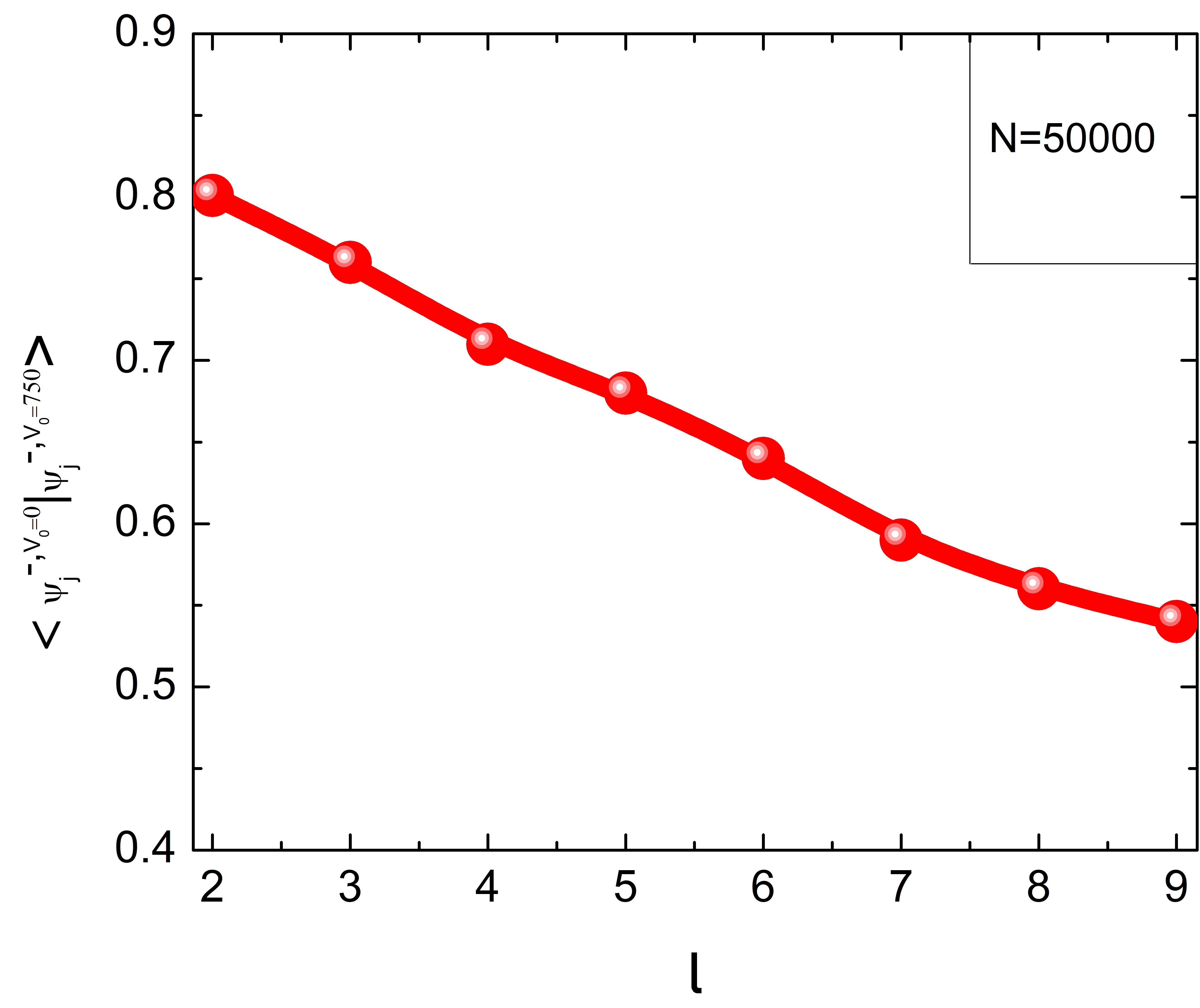}
     \caption{For larger values of $l$, the overlap of $\psi_j^{-,V_0=0}$ with $\psi_j^{-,V_0=750}$ is decreasing, where $\psi_j^{-,V_0=0}$ and $\psi_j^{-,V_0=750}$ are the $\psi_j^{-}$ (see equation \ref{psi_j-}) for the Rydberg electron-atom interaction strength $V_0=0$ and $V_0=750$ respectively. This suggests that the surface energy of the droplet inside the electron of the Rydberg atom tends to zero as $k\rightarrow 0$ as the overlap of the excited state tends to unity.}
\end{figure}
\begin{figure}
      \includegraphics[width=0.50\textwidth]{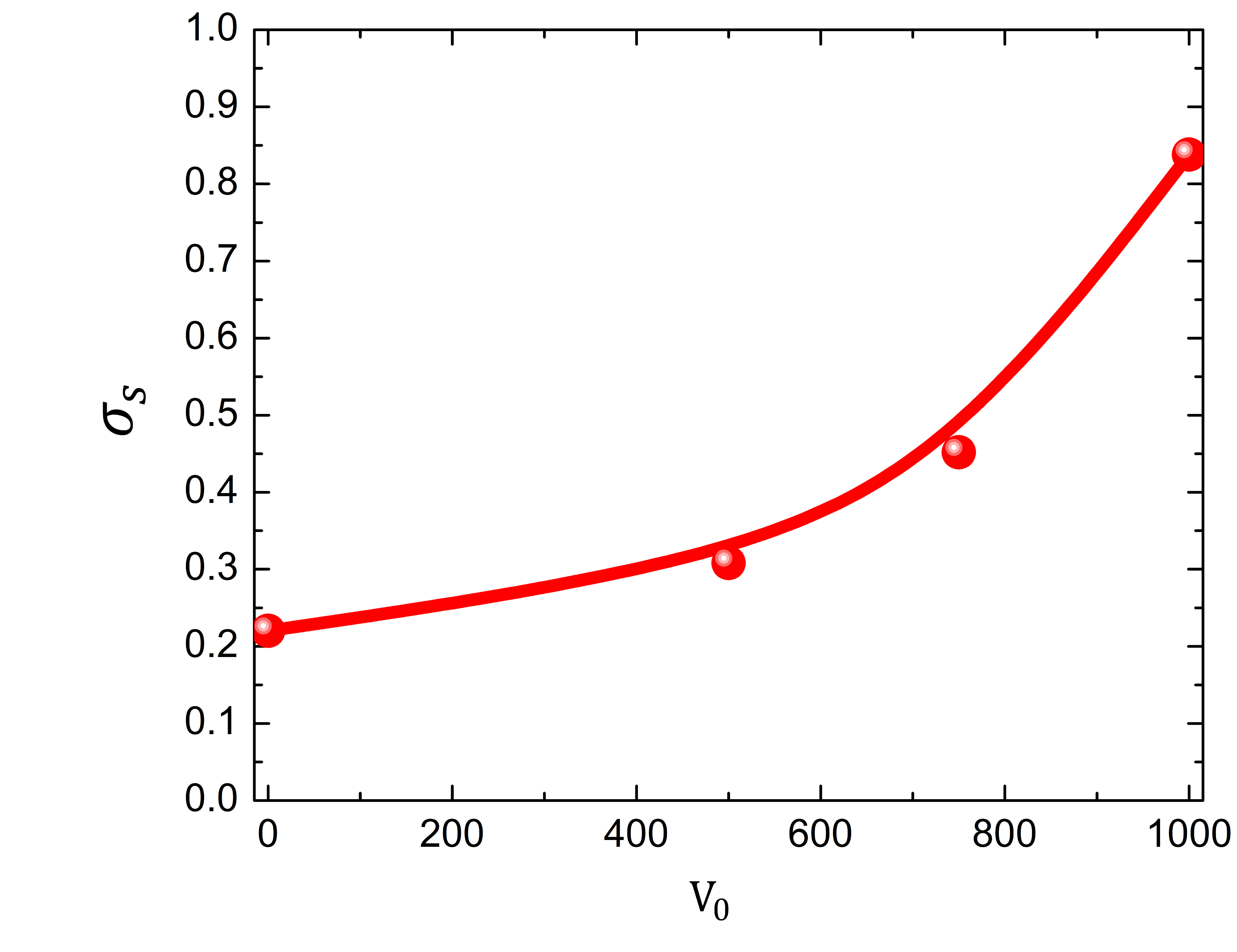}
     \caption{Surface tension ($\sigma_s)$ vs  Rydberg electron-atom interaction strength ($V_0$). We have fitted the surface excitation Figure 3 with expected dispersion relation (\ref{ripplons}) for surface modes  ($\omega_s= \sqrt{\sigma_s} k^{3/2}$)   and got the surface tension, which is increasing with $V_0$. Dots are the calculated values; line to track the tendency.}
\end{figure}


The expected dispersion relation between surface mode spectra ($\omega_s$) and the effective wave vector ($k$) is given by \cite{Petrov2015,landau}

\begin{equation}
\omega_{s} = \sqrt{\frac{4\pi l (l-1)(l+2)\sigma_s}{3}}N^{-1/2} = \sqrt{\sigma_s}k^{3/2}
\label{ripplons}
\end{equation}
Where $\sigma_s$ is the surface tension of the droplet and the effective wave vector $k$ is defined as $k=[l(l-1)(l+2)]^{1/3}/R$, $R$ is the size of the droplet. This $k^{3/2}$ fitting is shown in Fig. 3 by the black solid line.

\section*{Results and discussion}
Figure 1 is obtained by solving the GP equation (\ref{single_mode}) by the imaginary-time split-step Crank-Nicolson method. We chose a number $N$ and solved the GP equation (3) to check whether the volume is completely filled or not. The volume enclosed by the electron cloud is proportional to $n_r^6$, where $n_r$ is the principal quantum number of the Rydberg electron. The total number of condensed atoms ($N$) inside the electron of the Rydberg atom should depend on $n_r$ and the interaction between condensed atoms and Rydberg electron $(V_0)$.
We have calculated the maximum number of atoms enclosed by the Rydberg electron for different values of $n_r$ and the interaction strength between electron and atoms ($V_0$). We have plotted the $N$ as a function of $n^6_r$ in Figure 1 (i.e., the maximum number of atoms ($N$) can be fitted inside the Rydberg electron). The density of the condensate increases with $V_0$ (when the electron-atom interaction is strong, a larger number of particles can be fitted inside the Rydberg electron). The plot is a straight line, and the density of the droplet is the same for each line (for a given $V_0$); the only difference is that the radial extension ($R$) of the droplet increases with the larger combination of  $N$ and $n_r$. The collective excitations are independent of $n_r$ as the number of particles ($N$) is chosen from Figure 1. These help us to get different effective wave vectors ($k$).  Different values of $k$ can be obtained in two ways: (i) by changing angular momentum $l$ or (ii) by changing radial extension $R$ \cite{Jain}. But we can get a limited number of points if we change $l$ (though that is enough to study nature) for a given $N$. That's why, to get more data points (Fig. 3), we have varied $N$ according to Figure 1.  

We have plotted some of the excited state wave functions in Fig. 2. In the upper panel of the figure, i.e., Fig. 2[a], we have plotted the wavefunctions for a free droplet of $N=50000$. The "black dash-dot curve" is the ground state density obtained by solving equation (\ref{single_mode}), and the variations of $u_j$ and $v_j$ are shown for different values of angular momentum $l$. Fig. 2 [b] shows the wavefunctions for the droplet inside the Rydberg electron. Ground state density is increased due to electron-atom interactions, and variations of $u_j$ and $v_j$ are shown for different electron-atom interactions ($V_0$). The excitation wave function $u_j$ has no node in the radial direction as it is the wave function corresponding to the surface mode, which is located around the surface of the droplet.

The surface excitation energy spectrum has been represented in Fig. 3.
To calculate the energy spectrum, we have used a completely filled Rydberg atom; the number of particles ($N$) and $n_r$ have been chosen from Fig. 1. The black solid line is the $k^{3/2}$ (see equation \ref{ripplons}) fitting function in the low momentum region (up to k = 0.4) for the free droplet, which passes through the origin.
The surface mode spectra of the Rydberg-dressed droplet have a higher value than the free droplet. At the long wavelength limit, the surface mode for the Rydberg-dressed droplet will touch the origin with the surface mode without the Rydberg-dressed droplet, which is confirmed by the calculation of overlaps of the excited wave function of the free and Rydberg atom-dressed droplet (see figure 4).
The curve of the data points (Fig. 3) abruptly becomes flat for greater effective wave vectors (that is why we have used the lower wave-vector region for calculations of surface tension.). The particle-emission threshold |$\mu$| starts to merge with the surface mode frequency, at which point the mode frequency $\omega_{l0}$ can no longer be represented by the ripplon dispersion (\ref{ripplons}) (expected spectrum of the droplet’s surface mode) \cite{Petrov2015}. This is brought on by either a limited number of particles or a large angular momentum, $l$. Experimentally, for various high numbers of particles, the frequency of the quadrupole surface mode $\omega_{20}$ could be easily determined, together with the radius R of the droplet. The dispersion relation could then be confirmed, and the surface tension $\sigma_s$ could subsequently be measured empirically. The excitation curve (Figure 3) was then fitted using $ \omega= \sqrt{\sigma_s} k^{3/2} $ to get the surface tension ($\sigma_s$) of the stable droplet for different values of Rydberg electron-atom interaction. The surface tension has been plotted as a function of electron-atom interaction in Figure 5.  We have seen that the surface tension ($\sigma_s$) increases with the value of the electron-atom interaction.  Surface tension obtained from the surface excitation curve agrees with the calculated surface tension using the direct formula given in ref. \cite{Petrov2015}.
In conclusion, we have observed that the droplet becomes more stable than a free droplet inside the electron of the Rydberg atom as the surface tension increases with the Rydberg electron-atom interactions.

\end{document}